\documentclass[noshowpacs,preprint,aps,pre]{revtex4}
\usepackage[dvips]{graphicx}
\usepackage{amssymb}
\usepackage{amsmath}
\usepackage{latexsym}
\usepackage{epsfig}
\usepackage{bm}
\usepackage{color}
\usepackage{times,psfrag,subfigure}     
\usepackage{url}   
\begin{document}
\title{Defect Motifs for Constant Mean Curvature Surfaces}
\author{Halim Kusumaatmaja and David J. Wales}
\affiliation{
University Chemical Laboratories, University of Cambridge, Lensfield Road, Cambridge CB2 1EW, U.K.}
\date{\today}


\begin{abstract}
The energy landscapes of electrostatically charged particles embedded on
constant mean curvature surfaces are analysed for a wide range of system size,
curvature, and interaction potentials. The surfaces are taken to be
rigid, and the basin-hopping method is used to locate the putative global
minimum structures. The
defect motifs favoured by potential energy agree with experimental observations
for colloidal systems:
extended defects (scars and pleats) for weakly positive and negative Gaussian
curvatures, and isolated defects for strongly negative Gaussian curvatures.
Near the phase boundary between these regimes the two motifs are in strong
competition, as evidenced from the appearance of distinct funnels in the
potential energy landscape. We also report a novel defect motif consisting of
pentagon pairs. \end{abstract}
\maketitle

{\it{Introduction}} - The distribution of electrostatically 
charged particles on curved surfaces \cite{Thomson}
provides a testing ground for global optimisation algorithms
\cite{Wille86,Altschuler94,ErberH95,ErberH97,WalesU06,WalesMA09}, as well as
useful insights into a number of materials science and biological applications,
including the packing of virus capsids \cite{CasparK62,MarzecD93}, fullerene
structures \cite{krotohocs85,PerezGarrido}, colloidal crystals
\cite{BauschBCDHNNTW03,LipowskyBMNB05}, and proteins on lipid membranes
\cite{FresePOBWACHFG08}. A key issue in understanding the resulting structures
is the appearance of defects, which
strongly influence the mechanical, electronic and optical
properties. Defect structures have even been exploited for rational
design of templated self-assembly; for example, polar defects have been used to
create divalent metal nanoparticles \cite{DeVriesBHJLNUWS07}.

Frustration
is intrinsic to curved surfaces because of the competition between local order and
long-range geometrical constraints, and defects appear to screen the resulting
geometrical stresses, even for the ground state configuration. 
For example, while a hexagonal lattice is
possible for a flat surface, this arrangement cannot occur on a
sphere without introducing isolated five-fold disclinations (pentagons; with
positive topological charges), or composite structures with seven-fold
disclinations (heptagons; negative topological charges)
\cite{Wille86,Altschuler94,ErberH95,ErberH97,BauschBCDHNNTW03,WalesU06,WalesMA09,BowickCNT02,BowickG09}.
Generally, adjacent combinations of pentagons and heptagons may appear as
topologically uncharged or charged lines of dislocations, corresponding to pleats (e.g.~pentagon-heptagon 
topological dipoles) 
and scars (e.g.~pentagon-heptagon-pentagon
arrangements), respectively. 

Isolated heptagons may also exist if the Gaussian curvature of the substrate is
strongly negative, as was shown recently in studies of 
two-dimensional colloidal crystals on the
surface of capillary bridges \cite{IrvineVC10}. By systematically varying the shape and thus the
curvature of the substrate, a sequence of transitions was observed from zero
defects to isolated dislocations, pleats, scars, and isolated disclinations.

In the present contribution we characterise the energy landscape of electrostatically charged particles embedded
on constant mean curvature surfaces, modelling in particular the recent
experimental setup by Irvine et al.~for colloidal crystals
\cite{IrvineVC10,IrvineBC12}. Employing the basin-hopping algorithm
\cite{lis87, walesd97a,waless99,Wales03}, we identify likely global minimum
configurations for a wide range of system sizes and surface curvatures,
considering several different interaction potentials. Not only are we able to
reproduce the experimental sequence of defect transitions, but we also identify
a new defect motif corresponding to pentagon pairs, which may appear on
surfaces with both positive and negative Gaussian curvatures. Furthermore, we
show here for the first time the hierarchical nature of the potential energy
landscape for these systems. Using disconnectivity graphs \cite{BeckerK03}, we
demonstrate the appearance of separate funnels in the landscape, corresponding to
competing defect motifs favouring dislocations and disclinations. Our analysis
also provides information on the rearrangement mechanisms between defect
patterns, as well as insight into likely thermodynamic signatures for
structural transitions. 

{\it{Methodology}} - We consider $N$ identical electrostatically charged particles embedded on
the surface of catenoids (zero mean curvature) and unduloids (non-zero mean
curvature) \cite{Delauney}. To represent the interactions of these colloids
trapped on fluid interfaces \cite{LeunissenBHSC07}, we have mainly considered
Yukawa potentials of the form: $V = \sum^N_{j> i} \frac{1}{r_{ij}}
\,\, e^{-(r_{ij}/\lambda)}$, where $r_{ij}$ is the Euclidean distance between
particle $i$ and $j$, and $\lambda$ is the screening length. We choose $\lambda =
0.1$ so that the ratio between the screening length and the length of the
capillary bridges is of the same order as that used in experiments
\cite{IrvineVC10,LeunissenBHSC07}. 

We have also considered four other potentials: Yukawa with $\lambda = 0.4$,
Coulomb, Lennard-Jones, and a repulsive Lennard-Jones form. Qualitatively, the
defect motifs are the same provided that the particle density per unit area is
sufficiently high, indicating that our defect analysis should have wide-ranging
applicability to curved surfaces. This regime corresponds to $N > 200$. If the
number of electrostatically charged particles is smaller, they tend to populate the boundaries when Coulomb
or other similarly long-range potentials are used; for the Lennard-Jones
potential, we observe patches of hexagonal lattice separated by large spaces.
We therefore focus on $200<N<600$ in the present work.

\begin{figure*}
\centering
\includegraphics[angle=0]{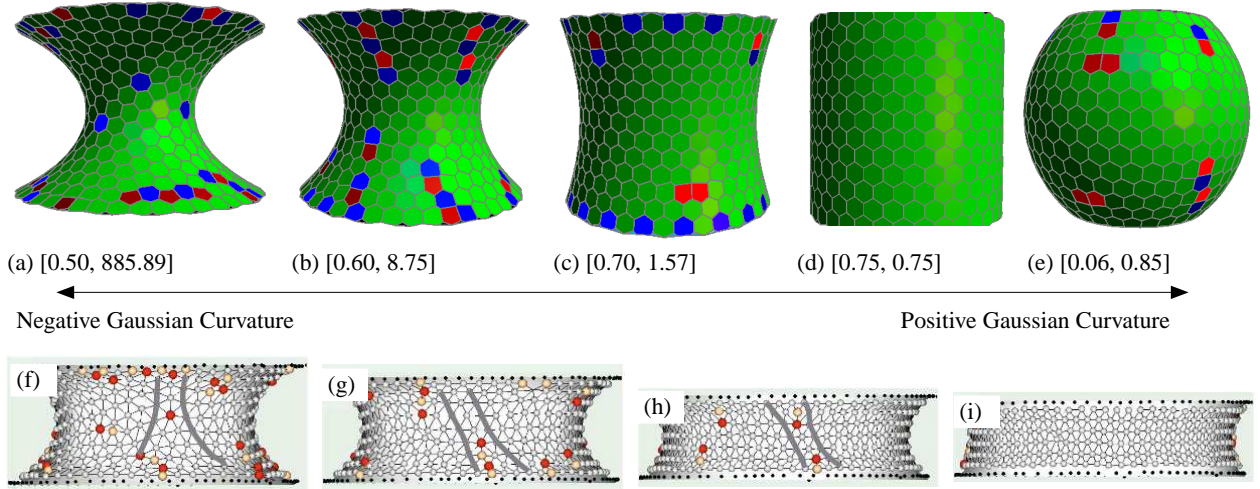}
\caption{Defect motifs for non-zero constant mean curvature surfaces. (a-e) The
number of electrostatically charged particles is $N=600$, and the curvature of the capillary bridges is
varied while keeping the height and volume of the bridges constant. The
corresponding values for the unduloid parameters [$a$, $c$] are given in 
square brackets. (f-i) Comparisons to experimental results of Irvine et al.~for
colloidal crystals, reproduced (will seek permission) from \cite{IrvineVC10}.
Here, red particles have seven neighbors, while yellow particles have five
neighbors. In the experiments, the curvature is tuned by stretching/compressing
the liquid bridge, and the sequence of defect transitions observed is identical to our
global optimisation results.} 
\label{Unduloid}
\end{figure*}

Any point on the surface of an unduloid or a catenoid can be parameterised
using two variables $u$ and $v$ \cite{HadzhilazovaMO07}. For catenoids, the
mapping to Cartesian coordinates takes the following form $\left(x,y,z \right)
= \left( c \cosh{(v/c)} \, \cos{(u)}, c \cosh{(v/c)} \, \sin{(u)}, v \right)$,
where $0 \leq u < 2\pi$, $-z_m \leq v \leq z_m$, and $c$ is a free parameter
corresponding to the waist of the catenoid in the $z=0$ plane. The
corresponding transformation is more elaborate for unduloids $\left( x, y, z
\right) = \left( (m\sin{\mu v}+n)^{1/2} \, \cos{(u)}, (m\sin{\mu v}+n)^{1/2} \,
\sin{(u)} , aF \left( \mu v/2-\pi/4,k \right) + c E \left( \mu v/2-\pi/4,k
\right) \right)$, with $\mu = 2/(a+c)$, $k^2=(c^2-a^2)/c^2$, $m=(c^2-a^2)/2$,
and $n=(c^2+a^2)/2$. $F(\phi,k)$ and $E(\phi,k)$ are elliptic integrals of the
first and second kind. We tune the shape of the unduloids by varying the
parameters $a$, $c$, and the range of values for $v$ \cite{HadzhilazovaMO07}.
For a neck shape unduloid [e.g. Fig. \ref{Unduloid}(a-c)], $v$ is centred
around $v_c=3\pi/2\mu$, while for a barrel shape unduloid, $v_c=\pi/2\mu$. The
maximum and minimum values of $v$ are chosen so that $-z_m \leq z \leq z_m$.

To characterise the most favourable geometries we employ basin-hopping global
optimisation \cite{lis87,walesd97a,waless99,Wales03}. In this method, random
geometrical perturbations are followed by energy minimisation, and moves are
accepted or rejected based upon the energy differences between local minima.
The minimisation procedure transforms the energy landscape of the system into
the set of catchment basins for the local minima, and makes the basin-hopping
method very effective for finding low-lying structures. 
Further details are provided in the supporting information.

We have also constructed databases of connected minima for selected systems,
starting from the low-lying structures found in the basin-hopping runs. We employed
doubly-nudged elastic band transition state searches \cite{TrygubenkoW04},
where images corresponding to local
maxima were tightly converged to transition states using hybrid
eigenvector-following \cite{HenkelmanUJ00,munrow99}. This procedure provides a
global survey of the potential energy landscape, which we then visualise using
disconnectivity graphs \cite{BeckerK03}. We find that distinct structural
arrangements of the particles result in separate funnels in the landscape.
Moreover, the database of connected minima and transition states allows us to
predict energy barriers and rearrangement mechanisms between defect rearrangements. 

To visualise the defect structures, we use Voronoi constructions. Thus,
pentagons, hexagons, and heptagons correspond to particles with five, six, and
seven neighbours. All the results presented here were obtained using the GMIN,
OPTIM, and PATHSAMPLE programs \cite{software}, which are available for use
under the GNU public license. 

{\it{Defect motifs on unduloids}} - We first analyse the arrangement of 
electrostatically charged particles
embedded on non-zero constant mean curvature surfaces (unduloids). The
results for $N=600$ identical particles are presented in Fig.~\ref{Unduloid}. The unduloid
parameters $a$ and $c$ were varied while keeping the height ($2\,z_m = 1.5$)
and volume ($V = 2.65$) of the unduloids constant. Experimentally, these parameters
correspond to capillary bridges for surfaces with different contact angles
\cite{KusumaatmajaL10,DeSouzaBMCA08}. The defect motifs and sequence of
transitions were found to be consistent as we varied the number of particles, and specific results
for $N=200$ are presented in the supporting information.

Due to the finite number of particles edge effects
are an intrinsic feature, and we find that the first two rows of Voronoi cells from the boundary
consistently have more defects. 
As for the experimental results on colloidal crystals
\cite{IrvineVC10} we find a series of
defect transitions, including isolated heptagons (Fig.~\ref{Unduloid}).
The Gaussian curvature is not constant on the surface of an unduloid, and it is
most negative at the waist, where isolated heptagons are located. Far from the
waist and near the edge of the unduloid, we observe lines of
dislocations.

For less negative total Gaussian curvatures [Figs \ref{Unduloid}(b-c)],
isolated heptagons disappear and dislocation lines prevail. The length of
the dislocations is also found to correlate strongly with the 
curvature: the length is shorter when the curvature is less negative. For
weakly negative Gaussian curvatures, the pentagon-heptagon dipole is a common
motif, together with a pair of pentagons. The pentagon pair motif is surrounded
by seven hexagons, as for an isolated heptagon, and to the best of our
knowledge, this motif has not been reported before. Since isolated pentagons
have a positive topological charge, it is somewhat surprising that double
pentagons may exist in a bound state, and that they can be favourable for
surfaces with negative Gaussian curvature.

We observe no defects for cylinders [zero Gaussian curvature,
Fig.~\ref{Unduloid}(d)], as expected. Dislocation lines as well as
pentagon pairs then reappear for unduloid with a
positive total Gaussian curvature [Fig.~\ref{Unduloid}(e)].
The orientations of the dislocations and
pentagon pairs are reversed for surfaces with positive and negative Gaussian
curvature. As predicted by continuum elastic theories
\cite{IrvineVC10,BowickY11}, for negative curvatures, the 7-5 dislocation axis
runs along the meridian, pointing in the direction of the boundary. For
positive Gaussian curvatures, the dislocation axis points to the centre of the
unduloid instead. Similarly, the pentagon pairs point to the boundary for
negative Gaussian curvatures, and to the centre of the unduloids for positive
Gaussian curvatures.  

\begin{figure*}
\centering
\includegraphics[angle=0]{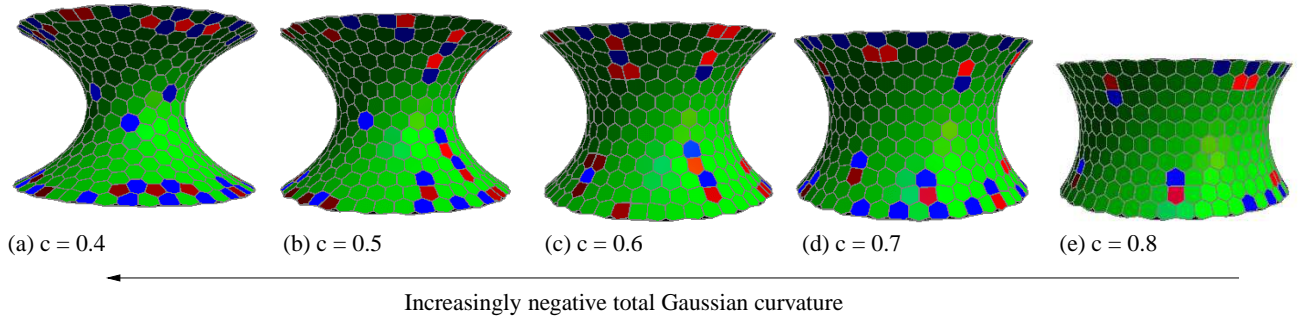}
\caption{Defect motifs on catenoid surfaces for $N=600$, with variable waist radius $c$.}
\label{Catenoid}
\end{figure*}

{\it{Defect motifs on catenoids}} - We have also analysed the defect structures for a family
of catenoids (minimal surfaces) with varying values of the waist parameter $c$.
We adjust the height parameter $z_m$ so that the radius of the catenoid at
$z_m$, $c \cosh{(z_m/c)}=1$. This procedure results in less negative total
Gaussian curvatures with increasing waist radius $c$, see Fig.~\ref{Catenoid}.

\begin{figure*}
\centering
\includegraphics[angle=0]{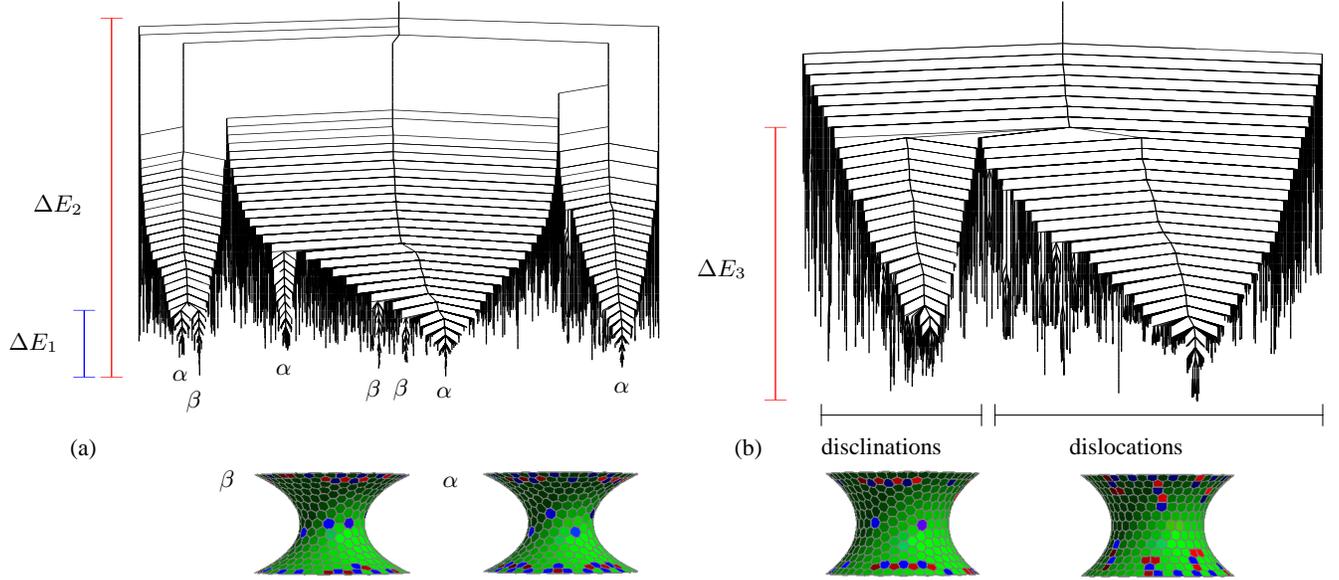}
\caption{Disconnectivity graphs for $N=600$ electrostatically charged particles embedded on catenoids with
waist radius (a) $c = 0.40$ and (b) $c=0.575$.
The insets in (a) and (b) show representative minima for the corresponding funnels in
the potential energy landscape. We also indicate the typical energy barriers
for interconverting different defect motifs. $\Delta E_1$, $\Delta E_2$, and $\Delta E_3$
are respectively 0.4, 2.2, and 0.7 in reduced units.} \label{DisGraph}
\end{figure*}

For strongly negative Gaussian curvature (small $c$ in Fig.~\ref{Catenoid}), we
always observe isolated heptagons at the waist of the catenoids, independent of
$N$. We illustrate the disconnectivity graph \cite{DisGraphExp} for $N=600$ and
$c=0.40$ in Fig.~\ref{DisGraph}(a). The potential energy landscape is clearly
hierarchical in character. Interestingly, there are only two dominant defect
configurations near the waist with a high degree of symmetry, which we label as
$\alpha$ and $\beta$ in the inset, each consisting of eight heptagons. The main
variation in the structural arrangements of the ions comes from the dislocation
lines near the edges (see the Supporting Information for animations). 
The energy barrier for interconverting waist
configurations $\alpha$ and $\beta$ is approximately $\Delta E_1 \sim 0.4$ (in
reduced units) if the configuration near the edge of the catenoid is
roughly preserved. On the other hand, the barrier for rearranging the particle
distribution near the edges can be much higher, $\Delta E_2 \sim 2.2$ (reduced
units), as indicated in Fig.~\ref{DisGraph}(a). From the pathway we see that
the high barrier is due to global concerted rotation of several layers of ions
near the edges of the catenoids. 

For weakly negative Gaussian curvatures (large $c$ in Fig.~\ref{Catenoid}), we
never find isolated heptagons, and only lines of dislocations and pentagon
pairs exist. It is worth noting that these defect motifs have the same
orientations as for unduloids with negative Gaussian curvature, 
above. The energy landscape for this parameter regime is also 
simpler. In particular, we find that the number of possible local minima is
considerably smaller. 

\begin{figure}
\centering
\includegraphics[angle=0]{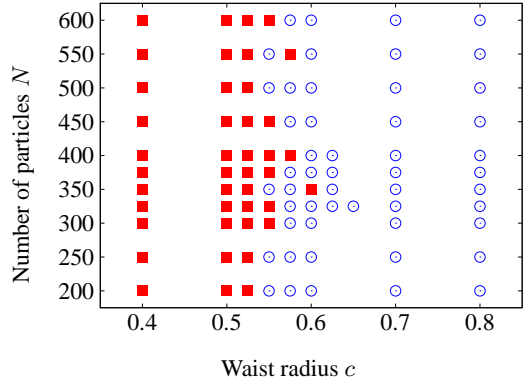}
\caption{Defect phase diagram as a function of the catenoid waist radius, $c$,
and number of electrostatically charged particles, $N$. Squares denote global minimum structures containing
isolated heptagons, while circles correspond to structures without isolated
heptagons.} \label{PhDiagram}
\end{figure}

The transition from defect patterns favouring disclinations to dislocations
occurs at moderately negative Gaussian curvatures, $c \sim 0.55$, which we
determine by constructing a defect phase diagram as a function of the number of
electrostatically charged particles, $N$, and the catenoid waist radius, $c$ 
(Fig.~\ref{PhDiagram}). Our global optimisation results further suggest that this
transition is only weakly dependant on the number of particles in this size
range. The phase diagram for the repulsive Lennard-Jones potential is shown
in the Supporting Information, where we find similar behaviour, except that the
phase boundary is shifted to larger $c$. Our results for
$N<600$ are probably far from the continuum limit, where Bowick and Yao
\cite{BowickY11} predict that $c$ will decrease monotonically with $N$.
Additionally, the phase boundary found in Fig.~\ref{PhDiagram} is neither
smooth nor monotonic. We can explain this observation by analysing the
disconnectivity graph for $N=600$ and $c=0.575$. As shown in
Fig.~\ref{DisGraph}(b), there are two distinct funnels, one favouring
dislocations and the other favouring disclinations. 
The energy difference between the lowest configurations in the two funnels 
is very small
compared to the energy barrier ($\Delta E_3 \sim 0.7$ in reduced units) 
associated with interconverting the two defect motifs.

{\it{Discussion}} - We have investigated the most favourable defect motifs for
electrostatically charged particles embedded on zero and non-zero constant mean curvature surfaces. By
varying the curvature of the embedding surface, we have characterised the
detailed structures and energetics of a wide range of defect motifs, including
dislocation lines and isolated disclinations, in excellent agreement with
experiment and predictions from continuum elastic theories. The appearance of a
new defect motif consisting of pentagon pairs is also predicted. 

Taking the typical experimental values for electrostatic interactions between
colloids embedded on interfaces, the energy scale for one simulation unit
corresponds to between 10 and $100k_B T$ (at room temperature). The energy
barriers of interest, as shown in Fig.~\ref{DisGraph}, are therefore quite
large and the structures observed in experiments could be trapped in
local minima. 
For experimentally relevant temperatures
the favoured defect motifs are primarily determined by the potential energy, 
which is consistent with the successful structure predictions that we have reported.
We hope these results will stimulate future theoretical and experimental work.
In particular, it will be interesting to modulate (reduce) the strength of
interactions between the electrostatically charged colloids, so that the defect rearrangement
mechanisms and thermodynamics of the system may be fully explored. Since the
energy landscape is hierarchical, with different energy scales for
interconverting the defect motifs, we expect the thermodynamics of the system
to be very rich. For example, we predict that there will be multiple
signatures in the heat capacity corresponding to alternative defect morphologies
\cite{Wales03}. Calculations of the thermodynamics, kinetic rates, and
rearrangement pathways for these defect motif transitions are currently
underway and will be presented elsewhere. 
The interplay between the arrangement of the electrostatically charged particles and possible deformation
of the interface is still an open question, and allowing for flexible rather than rigid
curved surfaces is another avenue for future research.

Analog models of the defect motifs presented here can be constructed using
Polydron tiles \cite{Polydron}, as illustrated in the Supporting Information (SI). 
The SI also contains animations of the defect morphologies shown in Figs
\ref{Unduloid}, \ref{Catenoid} and \ref{DisGraph}, the tabulated numbers of positive and negative topological charges, 
and detailed analysis of the net topological charges as a function of the integrated Gaussian curvature for several representative cases. 
These putative global minima found here will be made available online from the Cambridge Cluster Database \cite{CCD}.

{\it{Acknowledgements}} - This research was supported by EPSRC Programme grant EP/I001352/1
and by the European Research Council.

\bibliography{Thomson}

\begin{thebibliography}{37}
\bibitem[{Dis()}]{DisGraphExp}
\bibinfo{note}{In the disconnectivity graph, the vertical axis corresponds to
  the energy of the system. Each line ends with a local minimum. A node joins
  minima which can be interconverted without exceeding the energy of the node
  (thus providing information on the energy barriers between minima). For
  clarity, we discretise the energy in regular steps along the vertical axis.}
\end{thebibliography}

\end{document}